\pdfoutput=1

\PassOptionsToPackage{final}{hyperref}
\PassOptionsToPackage{final}{microtype}
\PassOptionsToPackage{hyphens}{url}

\RequirePackage{fancyvrb}
\RequirePackage[abort, l2tabu, orthodox]{nag}

\documentclass[acmsmall]{acmart}

\renewcommand\footnotetextcopyrightpermission[1]{} % removes footnote with conference information in first column
\usepackage[linesnumbered,ruled]{algorithm2e}
\usepackage{bookmark}
\usepackage{booktabs}
\usepackage{colortbl}
\usepackage{etoolbox}
\usepackage{flushend}
\usepackage[final]{listings}
\usepackage{newverbs}
\usepackage{pgfplots}
\usepackage{pgfplotstable}
\usepackage{subcaption}
\usepackage{tikz}

\usepackage{changes}
\definecolor{best}{gray}{.9}
\newcolumntype{b}{>{\columncolor{best}}r}

\newcommand{\subsubsubsection}[1]{\smallskip\noindent\textbf{#1.}}

\VerbatimFootnotes{}

\usetikzlibrary{
  arrows.meta,
  arrows,
  decorations.pathreplacing,
  positioning,
  quotes,
  scopes,
  shapes,
}

\usepackage{./sty/refs}
\usepackage{./sty/rqbox}
\usepackage{./sty/tikz-emph}
\usepackage{./sty/trfrac}

\lstdefinelanguage[word2vec]{C}{language=C, morekeywords={assume,call,Called,RetEq,RetNeq,RetLessThan,ParamTo,ParamShare,AccessPathStore,Sensitive,RetConst,PropRet,RetError,Error,FunctionStart,FunctionEnd,AccessPathSensitive}}

\crefname{enumi}{configuration}{configurations}
\creflabelformat{enumi}{#2(#1)#3}

\newcounter{columnsInGroup}

\pgfplotstableset{
  every head row/.append style={
    before row=\toprule,
    after row=\midrule,
  },
  every last row/.append style={
    after row=\bottomrule,
  },
  benchmarks/.append style={
    benchmark/.append style={
      assign column name/.append style={
        /pgfplots/table/column name/.add={\ttfamily}{},
      },
      score,
    },
    column group=Benchmark,
    columns/curl/.append style=benchmark,
    columns/hexchat/.append style=benchmark,
    columns/nginx/.append style=benchmark,
    columns/nmap/.append style=benchmark,
    columns/redis/.append style=benchmark,
    scores,
  },
  DAC improvements/.append style={
    benchmarks,
    every last row/.append style={
      before row=\midrule,
    },
    preproc cell content/.append code={
      \ifnumequal{\pgfplotstablerow + 1}{\pgfplotstablerows}{ % chktex 1
        \pgfkeys{/pgf/number format/showpos=true}
      }{}
    },
  },
  column group/.append style={
    every head row/.append style={
      before row/.append={
        \setcounter{columnsInGroup}{\pgfplotstablecols-1}
        & \multicolumn{\arabic{columnsInGroup}}{c}{#1} \\
        \cmidrule{2-\pgfplotstablecols}
      },
    },
  },
  highlight best/.append style={
    begin table/.add={
      \setlength{\extrarowheight}{\belowrulesep}
      \setlength{\aboverulesep}{0pt}
      \setlength{\belowrulesep}{0pt}
    }{},
    column group=Learner,
    columns/word2vec/.style=score,
    columns/GloVe/.style=score,
    columns/fastText/.style=score,
    every first row/.append style={
      before row=\rowcolor{best},
    },
    every last column/.append style={
      column type/.add={>{\columncolor{best}}}{},
    },
    scores,
  },
  scores/.append style={
    col sep=comma,
    column type=r,
    every first column/.append style={
      column type=l,
      string type,
    },
    score/.append style={
      fixed,
      fixed zerofill,
      multiply by=100,
      postproc cell content/.append style={
        @cell content/.add={}{\%},
      },
      precision=1,
    },
  },
  search path=fig,
}

\pgfplotsset{
  compat=1.9,
  varying alpha/.style={
    enlarge x limits=.15,
    grid=major,
    height=9cm,
    legend cell align=left,
    legend pos=#1 east,
    legend style={font=\ttfamily\vphantom{Xy}},
    xlabel=Alpha,
    xmax=1,
    xmin=0,
    xtick distance=.25,
    ylabel=Benchmark Score,
    ymax=1,
    ymin=0,
    xticklabel={\pgfmathprintnumber[fixed, fixed zerofill, precision=2]\tick},
    yticklabel={\pgfmathparse{\tick*100}\pgfmathprintnumber\pgfmathresult\%},
    /pgfplots/table/.cd,
    scores,
  }
}

\makeatletter
\renewcommand*{\NAT@spacechar}{~}
\makeatother

\title[Enabling Open-World Specification Mining]{Enabling Open-World Specification Mining via Unsupervised Learning}

\begin{document}

% \copyrightyear{2018} 
% \acmYear{2018} 
% \setcopyright{acmcopyright}
% \acmConference[ESEC/FSE 2019]{The 27th ACM Joint European Software Engineering Conference and Symposium on the Foundations of Software Engineering}{26--30 August, 2019}{Tallinn, Estonia}
% \acmBooktitle{Proceedings of the 26th ACM Joint European Software Engineering Conference and Symposium on the Foundations of Software Engineering (ESEC/FSE '18), November 4--9, 2018, Lake Buena Vista, FL, USA}
% \acmPrice{15.00}
% \acmDOI{10.1145/3236024.3236085}
% \acmISBN{978-1-4503-5573-5/18/11} % chktex 8

\newcommand{\UWMad}[1][ersity]{Univ#1 of Wisconsin--Madison} % chktex 8

\author{Jordan Henkel}
\orcid{0000-0003-3862-249X}
\affiliation[obeypunctuation=true]{%
  \institution{\UWMad},
  \country{USA}
}
\email{jjhenkel@cs.wisc.edu}

\author{Shuvendu K. Lahiri}
\affiliation[obeypunctuation=true]{%
  \institution{Microsoft Research},
  \country{USA}
}
\email{Shuvendu.Lahiri@microsoft.com}

\author{Ben Liblit}
\orcid{0000-0002-2245-2839}
\affiliation[obeypunctuation=true]{%
  \institution{\UWMad},
  \country{USA}
}
\email{liblit@cs.wisc.edu}

\author{Thomas Reps}
\orcid{0000-0002-5676-9949}
\affiliation[obeypunctuation=true]{%
  \institution{\UWMad[.] and GrammaTech, Inc.},
  \country{USA}
}
\email{reps@cs.wisc.edu}

\begin{abstract}
% 1. State the problem
% 2. Say why it's an interesting problem
Many programming tasks require using both domain-specific code and
well-established patterns (such as routines concerned with file IO). Together, several small patterns
combine to create complex interactions. This compounding effect, mixed with
domain-specific idiosyncrasies, creates a challenging environment for fully automatic
specification inference. Mining specifications in this environment, without the
aid of rule templates, user-directed feedback, or predefined API surfaces, is a
major challenge. We call this challenge Open-World Specification Mining.

% 3. Say what your solution achieves
In this paper, we present a framework for mining specifications and usage
patterns in an Open-World setting. We design this framework to be miner-agnostic
and instead focus on disentangling complex and noisy API interactions. To
evaluate our framework, we introduce a benchmark of 71 clusters extracted from
five open-source projects. Using this dataset, we show that interesting clusters
can be recovered, in a fully automatic way, by leveraging unsupervised learning
in the form of word embeddings. Once clusters have been recovered, the challenge
of Open-World Specification Mining is simplified and any trace-based mining
technique can be applied. In addition, we provide a comprehensive evaluation of
three word-vector learners to showcase the value of sub-word information for
embeddings learned in the software-engineering domain.

% 4. Say what follows from your solution

\end{abstract}

%%% cSpell:ignoreRegExp /\\cverb\|.*?\|/
%%% Local Variables: 
%%% mode: latex
%%% TeX-master: "paper.tex"
%%% End:

\maketitle
\thispagestyle{plain}

\renewcommand{\shortauthors}{J. Henkel, S. Lahiri, B. Liblit, and T. Reps}

\section{Introduction\label{Se:INTRO}}

% 1. Describe the problem
The continued growth of software in size, scale, scope, and complexity has
created an increased need for code reuse and encapsulation. To address this need,
a growing number of frameworks and libraries are being authored. These
frameworks and libraries make functionality available to downstream users
through Application Programming Interfaces (APIs). Although some APIs may be
simple, many APIs offer a large range of operations over complex structures
(such as the orchestration and management of hardware interfaces).

Staying within correct usage patterns can require domain-specific knowledge
about the API and its idiosyncratic behaviors \citep{robillard_field_2011}. This
burden is often worsened by insufficient documentation
and explanatory materials for a given API\@. In an effort to assist developers
utilizing these complex APIs, the research community has explored a wide variety
of techniques to automatically infer API properties
\citep{lo_mining_2011,robillard_automated_2013}.

This paper contributes to the study of API-usage mining by identifying a new
problem area and exploring the combination of machine learning and traditional
methodologies to address the novel challenges that arise in this new domain.
Specifically, we introduce the problem domain of Open-World Specification
Mining. The goal of Open-World Specification Mining can be stated as follows:

\begin{focusbox}
  Given noisy traces, from a mixed vocabulary, automatically identify and mine
  patterns or specifications without the aid of (i) implicit or explicit
  groupings of terms, (ii) pre-defined pattern templates, or (iii)
  user-directed feedback or intervention.   
\end{focusbox}

Open-World Specification Mining is motivated by the lack of adoption of
specification-mining tools outside of the research community. We believe that
because Open-World Specification Mining needs no user-supplied input, it will lead to tools
that are easier to transition and apply in industry settings. Although this setting
reduces the burden imposed on users, it increases the challenges
associated with extracting patterns. We address these challenges with a
toolchain, called \verb|ml4spec|:

\begin{itemize}
  \item
    We base our technique on a form of intraprocedural, parametric, lightweight symbolic execution
    introduced by~\citet{henkel_code_2018}. Using their tool gives us the
    ability to generate abstracted symbolic traces and avoids any need for
    dynamically running the program.
  \item
    To address the lack of implicit or explicit groupings of terms (a challenged
    imposed by the setting of Open-World Specification Mining) we introduce a
    technique, \emph{Domain-Adapted Clustering} (DAC), that is capable of recovering
    groupings of related terms.
  \item
    Finally, we remove the need for pre-defined pattern templates by mining specifications 
    using traditional, unrestricted, methods (such as k-Tails~\citep{biermann_synthesis_1972} and Hidden Markov Models~\citep{seymore_learning_1999}).
    We are able to use these traditional methods by leveraging Domain-Adapted
    Clustering to ``focus'' these traditional methods toward interesting
    patterns.
\end{itemize}

The combination of both traditional techniques and machine-learning-assisted
methods in the pursuit of Open-World Specification Mining raises a number of
natural research questions that we consider.

First, we explore the ability of Domain-Adapted Clustering, our key technique,
to successfully extract informative and useful clusters of API methods in our Open-World
setting:

\begin{rqbox}{canwe}
  Can we effectively mine useful and clean clusters of API methods in an Open-World setting?
\end{rqbox}

Immediately, we run into the difficulty of judging the utility of clusters
extracted from traces. To provide the basis for a consistent and quantitative
evaluation, we have manually extracted a dataset of ground-truth clusters from
five popular open-source projects written in C.

Next, we compare Domain-Adapted Clustering against several other baselines that
do not utilize the implicit structure of the extracted traces:

\begin{rqbox}{compare}
  How does Domain-Adapted Clustering (DAC) compare to off-the-shelf clustering
  techniques?
\end{rqbox}

We also explore how two key choices in our toolchain impact the
overall utility of our results:

\begin{rqbox}{choices}
  How does the choice of word-vector learner and the choice of sampling
  technique affect the resulting clusters?
\end{rqbox}

Understanding how different pieces of our toolchain interact provides the ground
work for understanding how traditional metrics (co-occurrence statistics)
interplay with our machine-learning-assisted additions (word embeddings). To
quantify the usefulness of unsupervised learning in our approach, and to validate our central
hypothesis, we ask:

\begin{rqbox}{benefit}
  Is there a benefit from using a combination of co-occurrence statistics and word
  embeddings?
\end{rqbox}

Finally, we can explore how faithful we are to one of the key tenets of Open-World
mining: the lack of user intervention. To do so, we must understand what
level of hyper-parameter tuning is required to achieve reasonable results:

\begin{rqbox}{transfer}
  Does our toolchain transfer to unseen projects with minimal reconfiguration?
\end{rqbox}

The contributions of our work can be summarized as follows:

\begin{itemize}
\item \textbf{We define the new problem domain} of Open-World Specification Mining.
Our motivation is to increase the adoption of specification-mining techniques
by reducing the burden imposed on users (at the cost of a more challenging mining task).

\item \textbf{We create a toolchain} based on the key insight that unsupervised
learning (specifically word embeddings) can be combined with traditional metrics
to enable automated mining in an Open-World setting. 

\item \textbf{We introduce a benchmark} of 71 ground-truth clusters extracted from
five open-source C projects.

\item \textbf{We report on several experiments:}
\begin{itemize}
  \item In \sectref{Rq-CanWe}, we use our toolchain to recover, on average, two thirds of the ground-truth clusters in our benchmark automatically.
  \item In \sectref{Rq-DAC}, we compare our Domain-Adapted Clustering technique to three
  off-the-shelf clustering algorithms; Domain-Adapted Clustering provides, on average, a 30\% performance
  increase relative to the best baseline.
  \item In \sectref{Rq-SAndL}, we confirm our intuition that sub-word information
  improves the quality of learned vectors in the software-engineering domain;
  we also confirm that our Diversity Sampling (\sectref{TS}) technique increases performance by solving the problem of \textit{prefix dominance}.
  \item In \sectref{Rq-Alpha}, we quantify the impacts of our machine-learning-assisted approach.
  \item In \sectref{Rq-Transfer}, we find that just two configurations can be automatically tested to achieve performance that is within 10\% of the best configuration.
\end{itemize}
\end{itemize}

\emph{Organization.} The remainder of the paper is organized as follows: \sectref{Overview}~provides an overview of the \verb|ml4spec| toolchain. \sectref{PLSE}~reviews
Parametric Lightweight Symbolic Execution. \sectref{TS}~describes Diversity Sampling. \sectref{DAC}~introduces
Domain-Adapted Clustering. \sectref{Mining}~describes trace projection and mining. \sectref{Evaluation}~provides an overview of our evaluation methodology.
\sectref{Rq-CanWe}-\sectref{Rq-Transfer}~address our five research questions. \sectref{Threats}~considers threats to the validity of
our approach. \sectref{RelatedWork}~discusses related work. \sectref{Conclusion}~concludes.

%%% cSpell:disable

\begin{figure}
\setlength{\belowcaptionskip}{-15pt}

\tikzset{%
base/.style = {rectangle, rounded corners, draw=black,
               minimum width=4cm, minimum height=1cm,
               text centered, font=\sffamily},
process/.style = {base, minimum width=2.5cm, fill=orange!15},
}

\begin{tikzpicture}[
    node distance=2.0cm,
    every node/.style={fill=white, font=\sffamily},
    align=left
]
    
\node (dProg) [process] {Input Programs};
\node (dTraces) [process, below of=dProg] {Traces};
\node (dThresTraces) [process, below of=dTraces] {Thresholded Traces};
\node (dSampTraces) [process, below of=dThresTraces] {Sampled Traces};
\node (dWordVectors) [process, below right of=dSampTraces] {Word Embeddings};
\node (dWordMatrix) [process, below of=dWordVectors] {Word Embedding\\ Matrix ($B$)};
\node (dCooccurMatrix) [process, left=0.5cm of dWordMatrix] {Co-occurrence\\ Matrix ($A$)};
\node (dCombinedMatrix) [process, below left of=dWordMatrix] {Combined Matrix: \\$\alpha A + (1 - \alpha ) B$ };
\node (dClusters) [process, below of=dCombinedMatrix] {Clusters};
\node (dProjTraces) [process, below of=dClusters] {Projected Traces};
\node (dFSA) [process, above right=-0.25cm and 1.5cm of dProjTraces] {FSAs};
\node (dPFSA) [process, below right=-0.25cm and 1.5cm of dProjTraces] {HMMs};

\draw[->]
    (dProg) -- (dTraces);
\draw[->]
    (dTraces) -- (dThresTraces);
\draw[->]
    (dThresTraces) -- (dSampTraces);
\draw[->]
    (dSampTraces) -- (dWordVectors);
\draw[->]
    (dWordVectors) -- (dWordMatrix);
\draw[->]
    (dSampTraces) -- (dCooccurMatrix);
\draw[->]
    (dWordMatrix) -- (dCombinedMatrix);
\draw[->]
    (dCooccurMatrix) -- (dCombinedMatrix);
\draw[->]
    (dCombinedMatrix) -- (dClusters);
\draw[->]
    (dClusters) -- (dProjTraces);
\draw[->]
    (dSampTraces) -| ++(-3.75,-3) |- (dProjTraces);
\draw[->]
    (dProjTraces) -- (dFSA);
\draw[->]
    (dProjTraces) --  (dPFSA);

\draw[decorate,decoration={brace,raise=6pt,amplitude=10pt}, thick]
    (4.0, 0.5)--(4.0, -3.0) node[right=0.75cm, midway]{\textbf{Phase I:} Parametric Lightweight \\Symbolic Execution};
\draw[decorate,decoration={brace,raise=6pt,amplitude=10pt}, thick]
    (4.0, -3.25)--(4.0, -6.5) node[right=0.75cm, midway]{\textbf{Phase II:} Thresholding \\ and Sampling}; 
\draw[decorate,decoration={brace,raise=6pt,amplitude=10pt}, thick]
    (4.0, -6.75)--(4.0, -8.5) node[right=0.75cm, midway]{\textbf{Phase III:} Unsupervised\\ Learning}; 
\draw[decorate,decoration={brace,raise=6pt,amplitude=10pt}, thick]
    (4.0, -8.75)--(4.0, -13.25) node[right=0.75cm, midway]{\textbf{Phase IV:} Domain-Adapted \\ Clustering};
\draw[decorate,decoration={brace,raise=6pt,amplitude=10pt}, thick]
    (5.5, -13.5)--(5.5, -16.25) node[right=0.75cm, midway]{\textbf{Phase V:} Mining};
\end{tikzpicture}
\caption{Overview of the ml4spec toolchain\label{Fi:Overview}}
\Description[]{}
\end{figure}
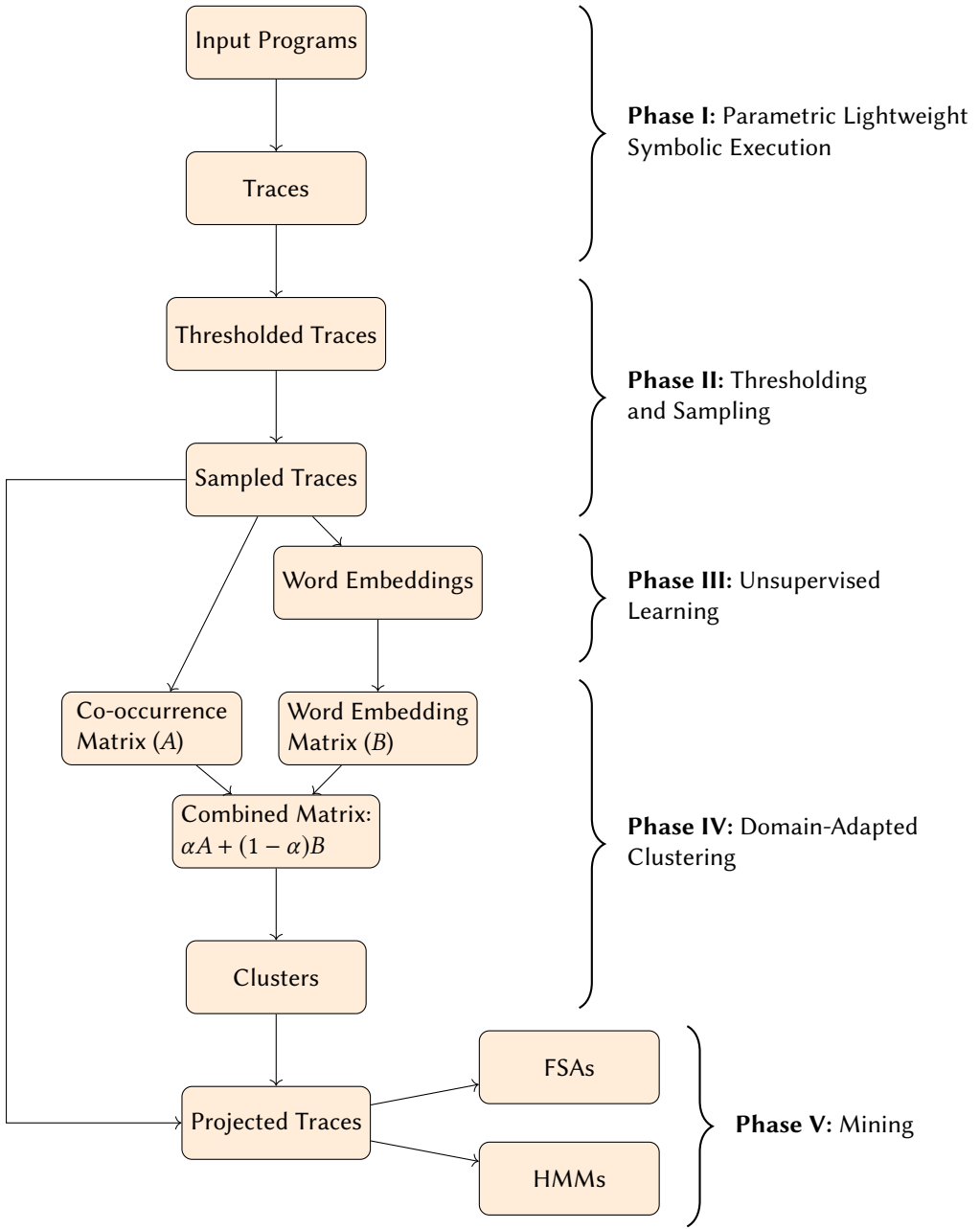

%%% cSpell:ignoreRegExp /\\cverb\|.*?\|/
%%% Local Variables:
%%% mode: latex
%%% TeX-master: "../paper.tex"
%%% End:

%%% cSpell:disable

\begin{figure}
\centering
\captionsetup{width=0.75\textwidth}
\begin{subfigure}{0.50\textwidth}
    \captionsetup{width=0.75\textwidth}
    \centering
\begin{lstlisting}
void example() {
  void *A,*B;
  if (!strcasecmp()) {
    addReplyHelp();
  } else if (!strcasecmp()) {
    A = dictGetIterator();
    log();
    while ((B = dictNext(A)) != 0) {
      dictGetKey(B);
      if (strmatchlen(sdslen())) {
        addReplyBulk();
      }
    }
    dictReleaseIterator(A);
  } else if (!strcasecmp()) {
    addReplyLongLong(listLength());
  } else {
    addReplySubcommandSyntaxError();
  }
}
\end{lstlisting}
    \caption{Sample procedure, showcasing an iterator usage pattern from the Redis open-source project\label{Fi:ExampleSrc}}
\end{subfigure}%
\begin{subfigure}{0.50\textwidth}
    \captionsetup{width=0.75\textwidth}
    \centering
\begin{lstlisting}[mathescape=true]
        $\$$START
        strcasecmp
        strcasecmp != 0
        strcasecmp
        strcasecmp == 0
        dictGetIterator
        log
        dictGetIterator $\rightarrow$ dictNext
        dictNext
        dictNext != 0
        dictNext $\rightarrow$ dictGetKey
        dictGetKey
        sdslen
        sdslen $\rightarrow$ strmatchlen
        strmatchlen
        strmatchlen != 0
        addReplyBulk
        dictGetIterator $\rightarrow$ dictNext
        dictNext
        dictNext == 0
        dictGetIterator
          $\rightarrow$ dictReleaseIterator
        dictReleaseIterator
        $\$$END
\end{lstlisting}
    \caption{One example trace, taken from the set of traces our Parametric Lightweight Symbolic Executor generates for the example procedure in~\figref{ExampleSrc}\label{Fi:ExampleTraces}}
\end{subfigure}
\caption{%
Example procedure and corresponding trace.
The notation $A \rightarrow B$ signifies that the result of call $A$ is used as a parameter to call $B$.\label{Fi:Example}}
\Description[]{}
\end{figure}

%%% cSpell:ignoreRegExp /\\cverb\|.*?\|/
%%% Local Variables:
%%% mode: latex
%%% TeX-master: "../paper.tex"
%%% End:

%% Local Variables:
%% mode: latex
%% TeX-master: "paper.tex"
%% End:

%%% cSpell:ignoreRegExp /\\cverb\|.*?\|/

% LocalWords:  pre DAC canwe dataset PLSE Rq CanWe RelatedWork SAndL

\section{Overview\label{Se:Overview}}

\newcommand{\headingtext}{}
\newcommand{\phasesubsection}[1]{%
  \refstepcounter{subsection}
  \renewcommand{\headingtext}{Phase \Roman{subsection}\texorpdfstring{\@}{}: #1}
  \subsubsubsection{\headingtext}
  \addcontentsline{toc}{subsection}{\headingtext}
}

The \verb|ml4spec| toolchain consists of five phases: Parametric Lightweight Symbolic Execution,
thresholding and sampling, unsupervised learning, Domain-Adapted Clustering, and mining. As input,
\verb|ml4spec| expect a corpus of buildable C projects. As output, \verb|ml4spec| produces clusters
of related terms and finite-state automata (or Hidden Markov Models) mined via traditional
techniques.\footnote{Although we provide examples based on traditional miners that produce finite-state automata and Hidden Markov Models, we do
note that the \verb|ml4spec| toolchain is miner-agnostic. By using \verb|ml4spec| as a trace pre-processor, any trace-based miner can be
adapted to the Open-World mining setting.} A visualization of the
way data flows through the \verb|ml4spec| toolchain is given in~\figref{Overview}. We illustrate
this process as applied to the example in~\figref{Example}.

\phasesubsection{Parametric Lightweight Symbolic Execution} The first phase of the \verb|ml4spec|
toolchain applies \emph{Parametric Lightweight Symbolic Execution} (PLSE). PLSE takes, as input, a corpus
of buildable C projects and a set of abstractions to apply. For our use case, we abstract calls,
checks on the results of calls, and return values. \sectref{PLSE} describes these abstractions
in more detail. Figure~\ref{Fi:Example} presents both an example procedure and a trace resulting
from the application of PLSE\@. Already, examining~\figref{ExampleTraces}, we can see one of the
core challenges of Open-World mining: the mixed vocabulary present in the trace from~\figref{ExampleTraces}
involves many interesting behaviors but, without user input~\citep{ammons_mining_2002,lo_smartic:_2006}, pre-defined rule templates~\citep{yun_apisan:_2016}, or some pre-described
notion of what methods are related~\citep{le_deep_2018}, we have no straightforward route to separating patterns from noise.
We need to \emph{disentangle} these disparate behaviors to facilitate better specification mining.

% FOOTNOTE about what the trace is saying? 

\phasesubsection{Thresholding and Sampling} Although our example procedure has a small number of paths
from entry to exit, many procedures have thousands of possible paths. Learning from these traces can
be challenging due to the number of times the same trace prefix is seen. This problem, which \citet{henkel_code_2018} term
\emph{prefix dominance}, makes downstream learning tasks more challenging. For instance, some terms
that occur in multiple traces (e.g., in a common prefix) may occur only a single time in the source program. Off-the-shelf word-vector
learners cannot filter for these kinds of rare words because they have no concept of the implicit hierarchy between
traces and the procedures they were extracted from. The \verb|ml4spec| toolchain introduces two novel
techniques to address these challenges: Diversity Sampling and Hierarchical Thresholding. Diversity Sampling
attempts to recover a fixed number of highly representative traces via a metric-guided sampling process.
Hierarchical Sampling leverages the implicit hierarchy between procedures and traces to remove rare words.
Together, these techniques improve the quality of downstream results. % TODO: Point to experiment

\phasesubsection{Unsupervised Learning} Traditionally, specification and usage mining techniques would
define some method of measuring support or confidence in a candidate pattern. Often, these measurements
would be based on co-occurrences of terms (or sets of terms). The \verb|ml4spec| toolchain leverages
a key insight: traditional co-occurrence statistics and machine-learning-assisted metrics (extracted via
unsupervised learning, specifically word embeddings) can be combined in fruitful ways.
Referencing our example in~\figref{Example}, we might hypothesize, based on co-occurrence, that \verb|dictGetIterator|
and \verb|log| are related. For the sake of argument, imagine that in each extracted trace we find this
same pattern. How can we refine our understanding of the relationship between \verb|dictGetIterator| and \verb|log|?

It is in these situations that adding unsupervised learning improves the results. A word-vector learner,
such as Facebook's fastText~\citep{bojanowski_enriching_2017}, can provide us with a measurement
of the similarity between \verb|dictGetIterator| and \verb|log|. This measurement provides a contrast to the
co-occurrence based view of our data. Intuitively, word-vector learners utilize the \emph{Distributional Hypothesis}:
similar words appear in similar contexts~\citep{harris_distributional_1954}. The global context, captured by co-occurrence statistics, can be supplemented
and refined by the local-context information that word-vector learners naturally encode. \sectref{Rq-Alpha} explores the
impact and relative importance of both traditional co-occurrence statistics and machine-learning-assisted metrics.

%%% cSpell:disable

\begin{figure}
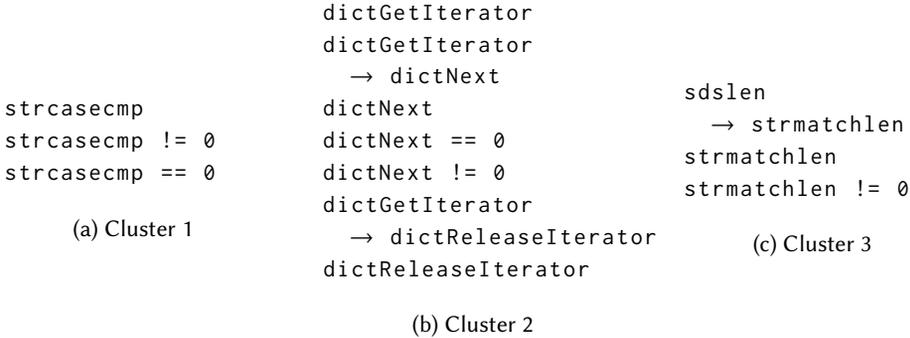

\centering
\begin{subfigure}{0.25\textwidth}
    \centering
    \begin{lstlisting}[mathescape=true]
strcasecmp
strcasecmp != 0
strcasecmp == 0
    \end{lstlisting}
    \caption{Cluster 1\label{Fi:Cluster1}}
\end{subfigure}%
\begin{subfigure}{0.40\textwidth}
    \centering
    \begin{lstlisting}[mathescape=true]
    dictGetIterator
    dictGetIterator
      $\rightarrow$ dictNext
    dictNext
    dictNext == 0
    dictNext != 0
    dictGetIterator
      $\rightarrow$ dictReleaseIterator
    dictReleaseIterator
    \end{lstlisting}
    \caption{Cluster 2\label{Fi:Cluster2}}
\end{subfigure}%
\begin{subfigure}{0.25\textwidth}
    \centering
    \begin{lstlisting}[mathescape=true]
sdslen
  $\rightarrow$ strmatchlen
strmatchlen
strmatchlen != 0
    \end{lstlisting}
    \caption{Cluster 3\label{Fi:Cluster3}}
\end{subfigure}
\caption{Clusters generated via Domain-Adapted Clustering (DAC)\label{Fi:Clusters}}
\Description[]{}
\end{figure}

%%% cSpell:ignoreRegExp /\\cverb\|.*?\|/
%%% Local Variables:
%%% mode: latex
%%% TeX-master: "../paper.tex"
%%% End:

% LocalWords:  DAC

%%% cSpell:disable

\begin{figure}
\begin{tikzpicture}[
    node distance=1.5cm,
    every node/.style={fill=white, font=\ttfamily},
    align=left
]
\node (0) [draw,ellipse] {~};
\node (1) [draw,ellipse,right of=0] {~};
\node (2) [draw,ellipse,right of=1] {~};
\node (3) [draw,ellipse,right of=2] {~};
\node (4) [draw,ellipse,right of=3] {~};
\node (5) [draw,ellipse,right of=4] {~};
\node (9) [draw,ellipse,below of=4] {~};
\node (6) [draw,ellipse,right of=5] {~};
\node (7) [draw,ellipse,right of=6] {~};
\node (8) [draw,ellipse,right of=7] {~};

\draw [->] (0) edge node[rotate=90,anchor=west]{\$START} (1);
\draw [->] (1) edge node[rotate=90,anchor=west]{dictGetIterator} (2);
\draw [->] (2) edge node[rotate=90,anchor=west]{dictGetIterator\\$\rightarrow$~dictNext} (3);
\draw [->] (3) edge node[rotate=90,anchor=west]{dictNext} (4);
\draw [->] (4) |- ++(-1.5,2.95) node[anchor=south]{dictNext != 0} -| (2);
\draw [->] (4) edge node[rotate=90,anchor=west]{dictNext == 0} (5);
\draw [->] (5) edge node[rotate=90,anchor=west]{dictGetIterator\\$\rightarrow$~dictReleaseIterator} (6);
\draw [->] (6) edge  node[rotate=90,anchor=west]{dictReleaseIterator}(7);
\draw [->] (7) edge node[rotate=90,anchor=west]{\$END} (8);
\draw [->] (3) edge node[anchor=east]{dictNext} (9);
\draw [->] (9) edge node[anchor=west]{dictNext == 0} (5);
\end{tikzpicture}
\captionsetup{width=0.75\textwidth}
\caption{Example FSA that was mined by projecting all of the traces extracted from~\figref{ExampleSrc} into the vocabulary defined by~\figref{Cluster2}. FSAs for the vocabularies defined by the clusters in~\figrefs{Cluster1}{Cluster3} are also generated, but not shown here.\label{Fi:DictIteratorFSA}}
\Description[]{}
\end{figure}
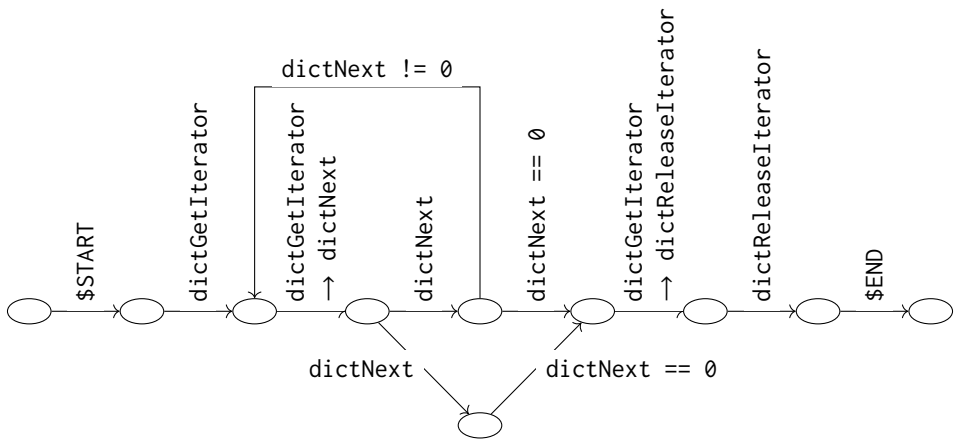

%%% cSpell:ignoreRegExp /\\cverb\|.*?\|/
%%% Local Variables:
%%% mode: latex
%%% TeX-master: "../paper.tex"
%%% End:

% LocalWords:  ExampleSrc Cluster2 FSAs Cluster1 Cluster3

\phasesubsection{Domain-Adapted Clustering} The trace given in~\figref{ExampleTraces} exhibits several different
patterns. The difficulty in mining from static traces like the one in~\figref{ExampleTraces} comes from the
need to learn a separation of the various, possibly interacting, patterns and behaviors. To address this challenge,
we introduce Domain-Adapted Clustering: a generalizable approach to clustering corpora of sequential data. Domain-Adapted Clustering
leverages the insight that it can be useful to combine machine-learning-assisted metrics with co-occurrence statistics
captured directly from the target corpus. Using Domain-Adapted Clustering, we can extract the clusters shown in~\figref{Clusters}.
These clusters allow us to solve the problem of \emph{disentanglement} by projecting the trace in~\figref{ExampleTraces}
into the vocabularies defined by each cluster. It is this ``focusing'' of the mining process that enables the \verb|ml4spec|
toolchain to apply traditional specification-mining techniques in an Open-World setting.

\begin{figure}
    \newlength{\nodedistance}
    \setlength{\nodedistance}{1.1cm}
    \begin{tikzpicture}[rotate=90, transform shape]
        \draw[every node/.style={
            align=center,
            draw,
            ellipse,
            font=\small\ttfamily,
            node distance=\nodedistance,
        }]
        node (0) {\$START}
        node (1) [below=of 0] {dictGetIterator}
        node (2) [below=of 1] {dictGetIterator \\$\rightarrow$~dictNext}
        node (3) [below=of 2] {dictNext}
        node (4) [below right=of 3] {dictNext\\!= 0}
        node (5) [below left=of 3] {dictNext\\== 0}
        node (6) [below=2\nodedistance of 3] {dictGetIterator \\$\rightarrow$~dictReleaseIterator}
        node (7) [below=of 6] {dictReleaseIterator}
        node (8) [below=of 7] {\$END}
        ;

        \draw
        [->, sloped, rotate=-90, every edge quotes/.style={anchor=south}]
        (0) edge["1.0"] (1)
        (1) edge["1.0"] (2)
        (2) edge["1.0"] (3)
        (3) edge["0.4"] (4)
        (4) edge["1.0", bend right] (2)
        (3) edge["0.6"] (5)
        (5) edge["1.0"] (6)
        (6) edge["1.0"](7)
        (7) edge["1.0"] (8)
        ;
    \end{tikzpicture}
\captionsetup{width=0.75\textwidth}
\caption{Example HMM that was mined by projecting all of the traces extracted from~\figref{ExampleSrc} into the vocabulary defined by~\figref{Cluster2}\label{Fi:DictIteratorHMM}}
\Description[]{}
\end{figure}
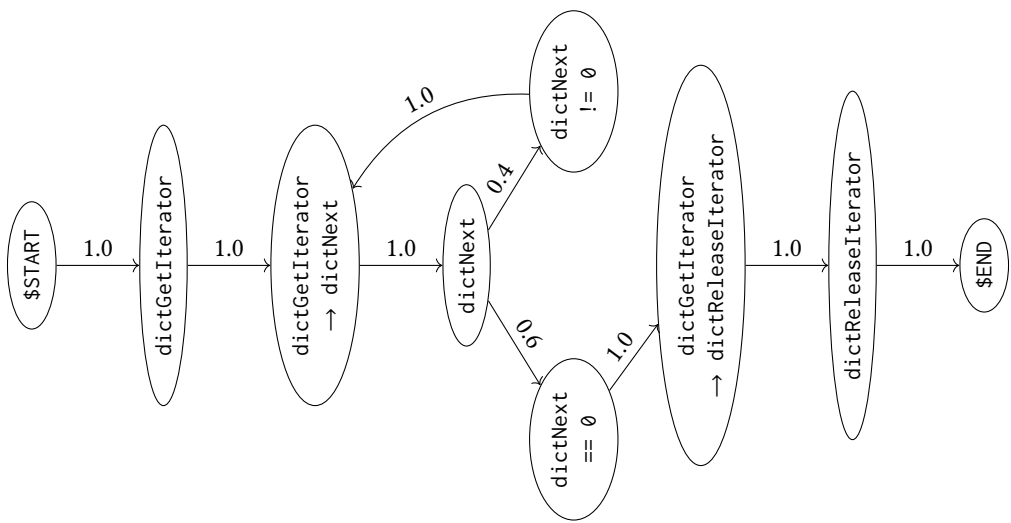

\phasesubsection{Mining} Finally, we can extract free-form specifications by applying traditional mining techniques
to the projected traces that \verb|ml4spec| creates. One powerful aspect of the \verb|ml4spec| toolchain is its disassociation
from any particular mining strategy. The real challenge of Open-World Specification Mining is extracting, without user-directed
feedback, reasonable clusters of possibly related terms. With this information in hand, a myriad of trace-based miners
can be applied. Figures~\ref{Fi:DictIteratorFSA} and~\ref{Fi:DictIteratorHMM} highlight this ability by showing both a finite-state automaton (FSA)
mined via the classic k-Tails algorithm and a Hidden Markov Model (HMM) learned directly from the projected traces~\citep{biermann_synthesis_1972,seymore_learning_1999}.

%%% cSpell:disable

%%% cSpell:ignoreRegExp /\\cverb\|.*?\|/
%%% Local Variables:
%%% mode: latex
%%% LaTeX-indent-level: 4
%%% TeX-brace-indent-level: 4
%%% TeX-master: "../paper.tex"
%%% End:

% LocalWords:  ExampleSrc Cluster2

\section{Parametric Lightweight Symbolic Execution\label{Se:PLSE}}

The first phase of the \verb|ml4spec| toolchain generates
intraprocedural traces using a form of parametric lightweight
symbolic execution, introduced by~\citet{henkel_code_2018}. Parametric
Lightweight Symbolic Execution (PLSE) takes, as input, a buildable C project
and a set of abstractions. Abstractions are used to parameterize the
resulting traces. In our setting, the abstractions allow us to
enrich the output vocabulary. This enrichment enables the final phase
of the \verb|ml4spec| toolchain (mining) to extract specifications
that include each of the following types of information:

\begin{itemize}
    \item \textbf{Temporal properties:} the \verb|ml4spec| toolchain abstracts
      the sequence of calls encountered on a given path of execution. The temporal
      ordering of these calls is preserved in the output traces.
    \item \textbf{Call-return constraints:} often a sequence of API calls can only
      continue if previous calls succeeded. In C APIs checking for success involves
      examining the return value of calls. \verb|ml4spec| abstracts simple checks over
      return values to capture specifications that involve
      call-return checks. % TODO: citation for parametric miners
    \item \textbf{Dataflow properties:} some specification miners are \emph{parametric}---these
      miners can capture relationships between parameters to calls and call-returns. To
      highlight the flexibility that PLSE provides, we include an abstraction that tracks
      which call results are used, as parameters, in future calls. This call-to-call dataflow
      occurs in many API usage patterns.
    \item \textbf{Result propagation:} the return value of a given procedure can encode valuable
      information. Some procedures act as wrappers around lower-level APIs, while other procedures
      may forward error results from failing calls. In either case, forwarding the result of a call,
      for any purpose, is abstracted into our traces to aid in downstream specification mining. 
      \verb|ml4spec| also abstracts constant return values: returning a constant
      may indicate success or failure, and such information may aid in downstream specification mining. 
\end{itemize}

With these various abstractions parameterizing our trace generation, simple downstream
miners, such a k-Tails, are capable of mining rich specifications. However, there is
a cost to the variety of abstractions we employ. Each abstraction introduces more words
into the overall vocabulary, and, as the size of the overall vocabulary grows, so does the
challenge of disentangling traces. 

Finally, it is worthwhile to address the limitations of Parametric Lightweight Symbolic
Execution. PLSE is intraprocedural and therefore risks extracting only
partial specifications. PLSE also makes no attempt to detect infeasible traces. Finally,
PLSE enumerates a fixed number of paths. As part of this enumeration, any loops are unrolled
for a single iteration only.\footnote{This single iteration loop unrolling gives us traces in which
the loop never occurred and traces in which we visit the loop body exactly one time. \Citet{yun_apisan:_2016}~follow a
similar model and argue that most API usage patterns are captured in a single loop unrolling.} In practice these limitations enable the PLSE technique to
scale and, for the purposes of the \verb|ml4spec| toolchain, losses in precision are
balanced by the utilization of machine-learning-assisted metrics (which can tolerate
noisy data).

% LocalWords:  PLSE

\section{Thresholding and Sampling\label{Se:TS}}

In this section, we outline the techniques used in the \verb|ml4spec| toolchain
to take a corpus of traces, generated via Parametric Lightweight Symbolic
Execution (PLSE), and prepare them for word-vector learning and specification mining.
In particular, we present two key contributions, Hierarchical Thresholding and
Diversity Sampling, which improve the overall quality of our results. In
addition, we discuss alternative approaches.

\subsection{Hierarchical Thresholding}

When preparing data for a word-vector learner, it is common to select a
\emph{vocabulary minimum threshold}, which limits the words for which vectors
will be learned. Any word that appears fewer times than the threshold
is removed from the training corpus. Through this process extremely
rare words, which may be artifacts of data collection, typos, or domain-specific
jargon, are removed. In the domain of mining specifications, we have a similar
need. We would like to pre-select terms, from our overall vocabulary, that occur
enough times to be used as part of a pattern or specification. We
could simply set an appropriate vocabulary minimum threshold using our word-vector
learner of choice; however, this approach ignores a unique aspect of our
traces. The traces we have, which are used as input to both the word-vector learner and
specification miner, are intra-procedural traces extracted from a variety of
procedures. To select terms that occur frequently does not necessarily select
for terms that are used across a variety of procedures. Because our traces are
paths through a procedure, it is possible to have a frequently occurring
term (with respect to our traces) that only occurs in one procedure.
To achieve our desire for terms that are used in a variety of
diverse contexts, we developed a modified thresholding approach: \emph{Hierarchical Thresholding}.
Hierarchical Thresholding counts how often a term occurs across
procedures instead of traces. This simple technique, with its utilization of
the extra level of hierarchical information that exists in the traces, reduces
the possibility of selecting terms that are rare at the source-code level but
frequent in the trace corpus.

\subsection{Diversity Sampling}

The corpus of symbolic traces that we obtain, via lightweight symbolic
execution, can be a challenging artifact to learn from. The symbolic executor,
at execution time, builds an execution tree and it is from this tree that we
enumerate traces. Any attempt to learn
from such traces can be thought of as an attempt to indirectly learn from the
original execution trees. The gap between the tree representation and trace
representation introduces a challenge: terms that co-occur at the start of a
large procedure (with many branches) will be repeated hundreds of times in our
trace corpus. This prefix duplication, which~\citet{henkel_code_2018} term
\emph{prefix dominance}, adversely affects the quality of word embeddings
learned from traces.

As part of the \verb|ml4spec| toolchain, we introduce a novel trace-sampling
methodology, which seeks to resolve the impact of \emph{prefix dominance}. We
call this sampling methodology \emph{Diversity Sampling} because it samples a
diverse and representative set of traces by using a similarity metric to drive
the sample-selection process.

\algref{DS}~provides the details of our Diversity Sampling technique. Because we
work with intra-procedural traces, we can associate each trace with its source-code
procedure. Consequently, the sampling routine can sample maximally diverse
traces for each procedure independently. (For a simple reason, our algorithm treats
the trace corpus as a collection of sets: each set holds the intra-procedural
traces for one source procedure.) To begin Diversity Sampling, we
either return all traces (if the number of traces for a given procedure is less
than our sampling threshold), or we begin to iterate over the available traces
and make selections. At each step of the selection loop, on lines 8--19, we identify a trace that
has the maximum average Jaccard distance when measured against our previous
selections. Jaccard distance is a measure computed between sets and, in our
setting, we use the set of unique tokens in a given trace to compute Jaccard
Distances. We take the average Jaccard distance from the set of currently
sampled traces to ensure that each new selection differs from all of the
previously selected traces. Finally, when we have selected an appropriate number
of samples, we return them and proceed to process traces from the next procedure.

%%% cSpell:disable

\begin{algorithm}
    \SetKwData{Choices}{choices}
    \SetKwData{D}{$D$}
    \SetKwData{DStar}{$D^*$}
    \SetKwData{S}{S}
    \SetKwData{Outputs}{outputs}
    \SetKw{Continue}{continue}
    \SetKwInOut{Input}{input}\SetKwInOut{Output}{output}
    \Input{A trace corpus $\mathbb{TR}$}
    \Output{A down-sampled trace corpus}
    \BlankLine
    \Outputs$\leftarrow []$\;
    \For{T $\in \mathbb{TR}$}{
        \If{$|T| \leq \text{SAMPLES}\;$}{
            \Outputs = \Outputs $\cup\;T$\;
            \Continue\;
        }
        \BlankLine
        \Choices$\leftarrow T[0]$\;

        \While{$|\;$\Choices$| < \text{SAMPLES}\;$}{
            \DStar$ = 0.0$\;
            \S$ = $ null\;
            \For{$t \in T - \text{choices}\;$}{
                \D$ = \text{AverageJaccardDistance}(t, \text{choices})$\;
                \If{$D \geq D^*$}{
                    \S$ = t$\;
                    \DStar$ =\;$\D\;
                }
            }
            \Choices = \Choices $\cup\;S$
        }

        \Outputs = \Outputs $\;\cup\;$ \Choices\;
    }
    \Return{\Outputs}\;
    \caption{Diversity Sampling\label{Alg:DS}}
\end{algorithm}

%%% cSpell:ignoreRegExp /\\cverb\|.*?\|/
%%% Local Variables:
%%% mode: latex
%%% TeX-master: "../paper.tex"
%%% End:
% LocalWords:  DStar

\subsection{Alternative Samplers}

Although Diversity Sampling is rooted in the intuition of extracting the most
representative set of traces for each procedure, it may not make a difference in
the quality of downstream results. It is for this reason that we also consider,
in our \verb|ml4spec| toolchain, two alternative approaches to trace sampling:
no sampling and random sampling. We include the option of no sampling because
word-vector learners thrive on both the \emph{amount} and \emph{quality} of
data available. It is reasonable to ask whether the training data lost by
downsampling our trace corpus has enough negative impact to offset possible
gains. We also include random sampling as a third alternative; our motivation
for this inclusion is to assess the impact of our metric-guided selection.
\sectref{Rq-SAndL} evaluates the sampling strategies
discussed here.

% LocalWords:  Thresholding PLSE pre intra thresholding DS Rq SAndL

\section{Domain-Adapted Clustering\label{Se:DAC}}

\sectref{PLSE} outlined how \verb|ml4spec| makes use of 
Parametric Lightweight Symbolic Execution (PLSE) to generate rich traces. In 
\sectref{TS}, we presented innovations that improved the traces
generated by PLSE, and addressed some of the challenges associated with
learning from traces. In this section, we introduce Domain-Adapted Clustering,
our solution to the challenge of clustering related terms. We seek to cluster
related terms (words) to simplify the Open-World Specification Mining
task. Traditional specification miners often use either rule templates or
some form of user-directed input (the API surface of interest, or perhaps a
specific class or selection of classes from which specifications should be
mined). In our Open-World setting, none of this information is available. Therefore, we have
developed a methodology for extracting clusters of related terms that harnesses the
power of unsupervised learning (in the form of word embeddings). With these
clusters in hand, the task of mining specifications is greatly simplified.

\subsection{Motivation}
To motivate Domain-Adapted Clustering, it is revealing to consider the
relationships among the following ideas:
\begin{itemize}
    \item \textbf{Co-occurrence:} word--word co-occurrence can be a powerful indicator of
      some kind of relationship between words. Co-occurrence is, by its nature, a global
      property that can, optionally, be associated with a sense of direction 
      (word $A$ appears to the left/right of word $B$).
    \item \textbf{Analogy:} analogies are another way in which words can be related. The words
      that form an analogical relationship encode a kind of information that is subtly different
      from the information that co-occurrence provides. Given the analogy $A$ is to $B$ as $C$ is to $D$,
      one would find that $A$ and $B$ often co-occur, as do $C$ and $D$; however, there may be no
      strong relationship (in terms of co-occurrence) between $A$/$B$ and $C$/$D$.
    \item \textbf{Synonymy:} synonyms are, in some sense, encoding strictly local structure. Two
      synonymous words need not co-occur; instead, synonyms are understood through the concept of
      replaceability: if one can replace $A$ with $B$ then they are likely synonyms.
\end{itemize}

We can now attempt to codify which of these concepts are of value for Open-World Specification Mining.
To do so, we will introduce a simple thought experiment: consider an extremely simple
specification that consists of a call to \verb|foo| and a comparison of the result of this call to \verb|0|.
In our traces this pattern would manifest in one of two forms: (i) \verb|foo foo==0| or (ii) \verb|foo foo!=0|.
For the sake of our thought experiment, also assume that, by chance, \verb|print| follows \verb|foo|
in our traces 95\% of the time. What kinds of relationships do we need to use to extract the
cluster of terms: \verb|foo|, \verb|foo==0|, and \verb|foo!=0|? We could use co-occurrence, however using
co-occurrence will likely pick up on the uninformative fact that \verb|foo| frequently co-occurs with \verb|print|.
Furthermore, co-occurrence may struggle to pick up on the relationship between \verb|foo| and the check on its result:
because each check is encoded as a distinct word, neither check will co-occur with extremely high frequency.
We could, instead, use synonymy, but it is trivial to imagine words, such as \verb|malloc| and \verb|calloc|,
that are synonyms but not related in the sense of a usage pattern or specification.

It is the insufficiency of both co-occurrence and synonymy that forms the basis of Domain-Adapted Clustering.
Because neither metric covers all cases, Domain-Adapted Clustering forms a parameterized mix of two metrics: one based
on left and right co-occurrence, and another based on unsupervised learning. Because both of these metrics
encode a distance (or similarity) of some sort, Domain-Adapted Clustering can be thought of as computing
the pair-wise distance matrices and then mixing them via a parameter $\alpha \in [0,1]$. Figure~\ref{Fi:Overview}
provides a visual overview of the mixing process Domain-Adapted Clustering employs.

\subsection{The Co-occurrence Distance Matrix}

Domain-Adapted Clustering utilizes co-occurrence statistics extracted directly from the (sampled and thresholded)
trace corpus. To capture as much information as possible, Domain-Adapted Clustering walks each trace and computes, for each word
pair $(A, B)$, the number of times that $A$ follows $B$ and the number of times that $B$ follows $A$. These counts
are converted to percentages and these percentages represent a kind of similarity between $A$ and $B$. The higher
the percentages, the more often $A$ and $B$ co-occur. To turn the percentages into a distance, we subtract them from
$1.0$ and store the average of the left-distance and right-distance in our co-occurrence distance matrix.

\subsection{The Word-Embedding Distance Matrix}

To incorporate unsupervised learning, Domain-Adapted Clustering utilizes word-vector learners. The use of word-vector learners
in the software-engineering domain is not a new idea~\citep{henkel_code_2018,defreez_path-based_2018,ye_word_2016,nguyen_exploring_2017,pradel_deepbugs:_2018}. Many recent works have explored the power of embeddings in the
realm of understanding and improving software. What we contribute is, to the best of our knowledge, the first thorough
comparison of three of the most widely used word-vector learners in the application domain of software engineering.
We do this comprehensive evaluation to test an intuition that \emph{sub-word information improves the quality of embeddings
learned from software artifacts}. We base this intuition on the observation that similarly named methods have similar meaning.
\sectref{Rq-SAndL} provides the details of this evaluation.

Our choice of word-vector learners as an unsupervised learning methodology is a deliberate one. Earlier, we saw how synonymy
could be a useful (albeit incomplete) property to capture. Furthermore, we already have a notion of distance between words
(given to us via our co-occurrence distance matrix). Word-vector learners mesh well with both of these pre-existing criteria: word vectors
encode local context and are able to capture synonymy. Additionally, word--word distance is encoded in the learned vector space. These properties
make word-vector learners a convenient choice for Domain-Adapted Clustering. To extract a distance matrix from a learned word embedding,
Domain-Adapted Clustering computes, for each word pair $(A, B)$, the cosine distance between the embedding of $A$ and the embedding of $B$ (here, we use cosine distance
because it is the distance of choice for word vectors). 

\subsection{Cluster Generation}

To generate clusters, Domain-Adapted Clustering applies the insight that the
clusters we seek should be expressed in concrete usages. This idea leads us to invert the
problem of clustering---instead of clustering all of the terms in the vocabulary, we take a 
more bottom-up approach. We start with an individual trace from our corpus of sampled traces. Within 
the trace, we find topics or collections of terms that are related under our machine-learning-assisted
metric: we use the combined distance matrix we created previously and apply a
threshold to detect words that are related. Within a trace, any two words whose distance is below
the threshold are assigned to the same intra-trace cluster. The next step uses the set of all intra-trace clusters
to create a set of reduced traces: each trace in the corpus of traces is projected onto each of the
intra-trace clusters to create a new corpus of reduced traces. To form final clusters, we apply a
traditional clustering method (DBSCAN~\citep{Ester:1996:DAD:3001460.3001507}) to the collection of reduced traces. In this final step we 
use Jaccard distance between the sets of tokens in the reduced traces as the distance metric.

One distinctive advantage of Domain-Adapted Clustering, for our use case, is its ability to generate overlapping
clusters. Most off-the-shelf clustering techniques produce disjoint sets but, in the realm of Open-World
Specification Mining, it is easy to conceive of multiple patterns that share common terms (opening a file
and reading versus opening a file and writing). Finally, it is worthwhile to note
that the clustering step we have outlined here introduces two hyper-parameters: the threshold to use
for intra-trace clustering (which we will call $\beta$) and DBSCAN's $\epsilon$, which controls how close
points must be to be considered neighbors. This leaves Domain-Adapted Clustering with a total of three
tunable hyper-parameters: $\alpha$, $\beta$, $\epsilon$.

% To generate clusters Domain-Adapted Clustering feeds a mixture of
% both the co-occurrence distance matrix and the word embedding distance
% matrix to DBSCAN (an off-the-shelf clustering algorithm). DAC utilizes
% a hyper-parameter, $\alpha \in [0,1]$, to control the relative weight of co-occurrence distance
% to word embedding distance. In particular, if we call the co-occurrence distance
% matrix $A$ and the word embedding distance matrix $B$, DAC produces a combined
% distance matrix: $\alpha A + (1-\alpha) B$. This combined distance matrix is then
% given to DBSCAN to produce clusters. DAC employs DBSCAN for cluster generation 
% because of DBSCAN's unique ability to
% tolerate uneven cluster sizes. To use DBSCAN two hyper-parameters must be
% set: $\epsilon$, which controls how close points must be to be considered
% neighbors and \verb|minSamples|, which defines how many
% neighbors are required to build a cluster. An exploration of the impact of
% all three hyper-parameters ($\alpha$, which we require to mix metrics and $\epsilon$ and \verb|minSamples|, which DBSCAN requires)
% is provided in section~\sectref{Evaluation}. 

% LocalWords:  PLSE Rq SAndL pre intra DBSCAN DBSCAN's

%%% cSpell:disable

\begin{figure}
  \centering
  \begin{subfigure}{0.5\textwidth}
      \centering
\begin{lstlisting}[mathescape=true]
        $\$$START
        strcasecmp
        strcasecmp != 0
        strcasecmp
        strcasecmp == 0
        dictGetIterator
        log
        dictGetIterator
          $\rightarrow$ dictNext
        dictNext
        dictNext != 0
        dictNext $\rightarrow$ dictGetKey
        dictGetKey
        sdslen
        sdslen $\rightarrow$ strmatchlen
        strmatchlen
        strmatchlen != 0
        addReplyBulk
        dictGetIterator
          $\rightarrow$ dictNext
        dictNext
        dictNext == 0
        dictGetIterator
          $\rightarrow$ dictReleaseIterator
        dictReleaseIterator
        $\$$END
\end{lstlisting}
      \caption{Example trace\label{Fi:ProjOrig}}
  \end{subfigure}%
  \begin{subfigure}{0.5\textwidth}
  \begin{minipage}{\textwidth}
      \centering
\begin{lstlisting}[mathescape=true]
      dictGetIterator
      dictGetIterator
        $\rightarrow$ dictNext
      dictNext
      dictNext == 0
      dictNext != 0
      dictGetIterator
        $\rightarrow$ dictReleaseIterator
      dictReleaseIterator
\end{lstlisting}
      \caption{Example cluster\label{Fi:ProjCluster}}
  \end{minipage}
  \begin{minipage}{\textwidth}
    \captionsetup{width=0.75\textwidth}
    \centering
\begin{lstlisting}[mathescape=true]
      dictGetIterator
      dictGetIterator
        $\rightarrow$ dictNext
      dictNext
      dictNext != 0
      dictGetIterator
        $\rightarrow$ dictNext
      dictNext
      dictNext == 0
      dictGetIterator
        $\rightarrow$ dictReleaseIterator
      dictReleaseIterator
\end{lstlisting}
  \caption{Result of projecting the trace in~\figref{ProjOrig} into the vocabulary defined by the cluster in~\figref{ProjCluster}\label{Fi:ProjResult}}
  \end{minipage}
  \end{subfigure}
  \caption{Example of trace projection\label{Fi:Projection}}
  \Description[]{}
\end{figure}

%%% cSpell:ignoreRegExp /\\cverb\|.*?\|/
%%% Local Variables:
%%% mode: latex
%%% TeX-master: "../paper.tex"
%%% End:

% LocalWords:  ProjOrig ProjCluster

\section{Mining\label{Se:Mining}}

The final phase of the \verb|ml4spec| toolchain is mining. To mine
specifications in an Open-World setting, \verb|ml4spec| applies several
insights, described in the preceding sections, to create a corpus of rich traces and a collection of clusters.
These two artifacts are used, in the mining phase, to create a new corpus
of \emph{projected traces} that can be fed to any pre-existing trace-based
mining technique. To create projected traces, \verb|ml4spec| takes each cluster
and each trace and generates a new projected trace by removing, from the original
trace, any word that is not in the currently selected cluster. This projection
process is shown in~\figref{Projection}. After projection, traces can be de-duplicated (if desired)
and then passed to any trace-based miner. The \verb|ml4spec| toolchain is unique
in its non-reliance on any particular trace-based miner. It is the dissociation from
specific mining techniques that makes \verb|ml4spec| a toolchain for Open-World mining
and not just another trace-based mining technique.

% LocalWords:  pre de

\section{Evaluation\label{Se:Evaluation}}

In this section we introduce our evaluation methodology and address each
of our five research questions. For the purposes of evaluation we ran the \verb|ml4spec|
toolchain on five different open source projects:

\begin{itemize}
    \item \textbf{Curl:} a popular command-line tool for transferring data.
    \item \textbf{Hexchat:} an IRC client.
    \item \textbf{Ngnix:} a web server implementation.
    \item \textbf{Nmap:} a network scanner.
    \item \textbf{Redis:} a key-value store.
\end{itemize}

These projects were selected because they exhibit a wide variety of usage patterns
across diverse domains. For each of these five projects, we performed a grid search to gain an understanding
of our various design decisions. \sectref{GridSearch} details this search.

%%% cSpell:disable

\begin{table}
\caption{Grid search parameters\label{Ta:GridSearchParams}}
\Description[]{}
\pgfplotstabletypeset[
column type=l,
columns/Name/.style={
  postproc cell content/.style={
    @cell content/.add={\ttfamily}{},
  },
},
columns/Values/.style={
  postproc cell content/.style={
    @cell content/.add={\{}{\}},
  },
},
col sep=&,
row sep=\\,
string type,
]
{
Name    & Values                        & Purpose                              & Phase \\
learner & fastText, GloVe, word2vec     & Word-vector learner to use           & II (Learning) \\
sampler & Diversity, Random, None       & Sampling method to use               & III (Sampling) \\
alpha   & 0.00, 0.25, 0.50, 0.75, 1.00  & Weight for combined distance matrix  & IV (DAC) \\
beta    & 0.20, 0.25, \dots, 0.45, 0.50 & Threshold for intra-trace clustering & IV (DAC) \\
epsilon & 0.10, 0.30, 0.50, 0.70, 0.90  & Parameter to DBSCAN                  & IV (DAC) \\
}
\end{table}

%%% cSpell:ignoreRegExp /\\cverb\|.*?\|/
%%% Local Variables:
%%% mode: latex
%%% TeX-master: "../paper.tex"
%%% End:

% LocalWords:  fastText GloVe word2vec DAC intra DBSCAN

\subsection{Grid Search\label{Se:GridSearch}}

To facilitate a comprehensive evaluation of \verb|ml4spec|, we performed a grid search across thousands
of parameterizations of the \verb|ml4spec| toolchain. The grid search serves
two purposes. First, the results of the grid search provide a firmer empirical
footing for understanding the efficacy and impacts of different aspects of
our toolchain (in particular, the grid search aids in quantifying the impacts of different
word-vector learners and sampling methodologies). Second, the Open-World Specification Mining
task emphasizes a lack of user-directed feedback---to meet this standard we must
ensure, via the data gleaned from the grid search, that the hyper-parameters associated
with the \verb|ml4spec| toolchain can be set, globally, to good default values. Table~\ref{Ta:GridSearchParams}
outlines the parameters in play and the ranges of values investigated for each parameter. Upper and lower
limits for each search range were carefully chosen, based on the results of smaller searches, to reduce
the computational costs of the larger search over the parameters presented in~\tableref{GridSearchParams}.

\subsection{RQ1: Can we effectively mine useful and clean clusters in an Open-World setting?\label{Se:Rq-CanWe}}

\Cref{rq:canwe} asked whether we can mine useful and clean clusters. The
difficulty with mining such clusters lies in the setting of our mining task. We
seek to solve the problem of mining specifications in an Open-World setting: one
in which implicit and explicit sources of hierarchical or taxonomic information
are unavailable. It is this Open-World setting that creates a unique need for
disentangling the many different topics that may exist in an abstracted symbolic
trace. The purpose of \cref{rq:canwe} is to understand the efficacy of the
techniques described earlier (specifically, Domain-Adapted Clustering) against a
key challenge of Open-World Specification Mining: learning correct and clean
clusters.

To measure the quality of our learned clusters we found the need
for a benchmark. Unfortunately, to the best of our knowledge, the problem of
Open-World Specification Mining has not been previously addressed and,
therefore, there is no ground truth to evaluate our learned clusters against
(due to the lack of need for gold standard clusters). One possible avenue of
evaluation and source of implicit clusters exists in source-code documentation.
Many thoroughly documented and heavily used APIs include information on the
associations between functions (most commonly in the form of a ``See
also\ldots'' or ``Related methods\ldots'' listing). Another possible source of
implicit information comes from projects that have made the transition from a
language like C to a language like C++. In such a transition methods are often
grouped into classes and this signal could be used to induce a clustering.
Finally, there is the implicit clustering induced by the locations where various API
methods are defined: even in C, functions defined in the same
header are likely related.

%%% cSpell:disable

\begin{figure}
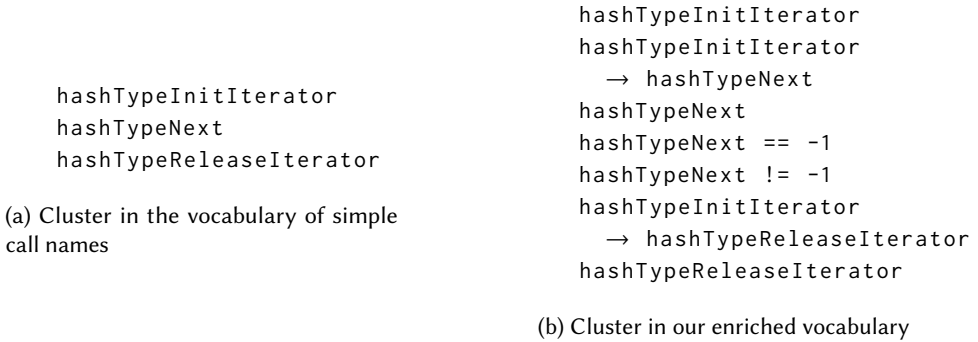

\centering
\captionsetup{width=0.75\textwidth}
\begin{subfigure}{0.50\textwidth}
    \captionsetup{width=0.75\textwidth}
    \centering
    \begin{lstlisting}[mathescape]
        hashTypeInitIterator
        hashTypeNext
        hashTypeReleaseIterator
    \end{lstlisting}
    \caption{Cluster in the vocabulary of simple call names\label{Fi:cluster-vocab-comp:simple}}
\end{subfigure}%
\begin{subfigure}{0.50\textwidth}
    \captionsetup{width=0.75\textwidth}
    \centering
    \begin{lstlisting}[mathescape]
        hashTypeInitIterator
        hashTypeInitIterator
          $\rightarrow$ hashTypeNext
        hashTypeNext
        hashTypeNext == -1
        hashTypeNext != -1
        hashTypeInitIterator
          $\rightarrow$ hashTypeReleaseIterator
        hashTypeReleaseIterator
    \end{lstlisting}
    \caption{Cluster in our enriched vocabulary\label{Fi:cluster-vocab-comp:enriched}}
\end{subfigure}
\caption{Comparison between two clusters\label{Fi:cluster-vocab-comp}}
\Description[]{}
\end{figure}

%%% cSpell:ignoreRegExp /\\cverb\|.*?\|/
%%% Local Variables:
%%% mode: latex
%%% TeX-master: "../paper.tex"
%%% End:

Despite these various sources of implicit clusters, we have identified a need
for manually defined gold standard clusters. We use manually extracted ground
truth clusters for two reasons. First, the sources of information listed above are
indications of relatedness but not necessarily indications of a specification or
usage pattern. For example, several different methods are commonly defined for
linked lists, such as \verb|length()|, \verb|next()|, \verb|prev()|, and
\verb|hasNext()| but not all of these methods are necessarily used together in a
pattern. Second, the vocabulary we are working over includes more than
simple call names---we also have information related to the path condition and
information about dataflow between calls. For example, compare the two
clusters given in~\figref{cluster-vocab-comp}. The cluster in~\figref{cluster-vocab-comp:simple}
consists only of call names, while the cluster in~\figref{cluster-vocab-comp:enriched} includes call
names, return value checks, and dataflow information. In comparing these two
clusters, it becomes clear that a cluster over words in our enriched vocabulary
(induced by the abstractions we choose) is strictly more informative than a
cluster over a vocabulary of simple call names. 

Taken together, these two issues (the weak signal of the aforementioned sources
and the lack of labels for some words in our enriched vocabulary) make manually
extracted clusters more desirable. For the purpose of this evaluation we have
extracted 71 gold standard clusters from five open source projects. We have placed
no explicit limit on the sizes of the clusters we included, thereby increasing the
challenge of recovering all the clusters in our benchmark correctly.

Using our set of 71 gold standard clusters we are able to perform a quantitative
evaluation by measuring the Jaccard similarity\footnote{Jaccard similarity between sets $A$ and $B$ is $\frac{|A \cap B|}{| A \cup B|}$.
Jaccard distance is one minus the Jaccard similarity.} of our extracted clusters and our
gold standard clusters. Because our toolchain does not mine a fixed number of
clusters, we need some way to ``pair'' an extracted cluster with the gold
standard cluster it most represents. To do this, we look for a pairing of
extracted clusters with gold standard clusters that maximizes the average
Jaccard similarity. This provides us with a way to have a consistent evaluation
regardless of the number of total clusters we extract. (One might argue that
this allows for extracting an unreasonable amount of clusters in an attempt to
game this metric.  However, this kind of ``metric hacking'' is unachievable in our
toolchain due to the use of DBSCAN to extract clusters from reduced traces.
Clustering our reduced traces, using the Jaccard distance between sets of tokens within a trace,
removes the possibility that our tool is simply enumerating all possible
clusters to achieve a high score.)

%%% cSpell:disable

\begin{table}
\captionsetup{width=0.75\textwidth}
\caption{Best scoring configurations for each of the five target projects\label{Ta:Best}}
\Description[]{}
\begin{filecontents*}{best-scoring-table.csv}
Measurement,curl,hexchat,nginx,nmap,redis
Top-1 (Jaccard),.628,.528,.449,.491,.719
Top-1 (Intersection),.797,.787,.676,.701,.835
Top-5 (Jaccard),.621,.518,.437,.473,.693
Top-5 (Intersection),.779,.738,.701,.672,.810
\end{filecontents*}
\pgfplotstabletypeset[benchmarks]{best-scoring-table.csv}
\end{table}

%%% Local Variables:
%%% mode: latex
%%% TeX-master: "../paper.tex"
%%% End:

In addition to Jaccard similarity, which penalizes both omissions and spurious inclusions, we
also measure the percent intersection between our extracted clusters and the clusters in our gold standard dataset. Table~\ref{Ta:Best}
provides both of these measurements for each of the five open-source projects we examined. To provide a robust
understanding of performance, and reduce variance in our measurements, we provide both the best (Top-1) results
and an average of the five best results (Top-5) for both similarity measures. (We use the data from our grid search,
described in~\sectref{GridSearch}, to compute these averages.) Examining~\tableref{Best},
we observe that the \verb|ml4spec| toolchain retrieves clusters that have a strong agreement
with the clusters in our gold standard dataset. Furthermore, the intersection similarity results show that our extracted
clusters contain, on average, over two thirds of the desired terms from the clusters in our gold standard dataset.
Together, these results answer \cref{rq:canwe} in the affirmative: \verb|ml4spec| is capable
of extracting clean and useful clusters in an Open-World setting.

\subsection{RQ2: How does DAC compare to off-the-shelf clustering techniques?\label{Se:Rq-DAC}}

In this section, we explore how our Domain-Adapted Clustering (DAC) technique (a
key piece of our Open-World specification miner) compares to traditional
clustering approaches. To understand the relationship between DAC and more
traditional clustering methods, it is instructive to consider the input data we
have available to use in the clustering process. Prior to clustering, we have
access to a pairwise distance matrix (created via a combination of
co-occurrence statistics and word--word cosine distance), our learned word
vectors, and our original traces.

Most clustering methods accept either vectors of data or pair-wise distance
matrices. In principle, this leaves our choices for clustering methods to
compare to quite open. However, using our word vectors as the input to
clustering ignores our earlier insight about the advantage of
combining word embeddings and co-occurrence statistics. Therefore, we focus
on clustering algorithms that accept pre-computed pair-wise distances as input.
From this class of clustering methods we have selected the following techniques
to compare against: DBSCAN~\citep{Ester:1996:DAD:3001460.3001507}, Affinity Propagation~\citep{Frey972}, and Agglomerative Clustering. 

\newcommand{\dacCompareTable}[3]{
  \begin{table}
    \captionsetup{width=0.75\textwidth}
    \caption{DAC compared to off-the-shelf clustering techniques #3\label{Ta:DACAlpha#1}}
    \Description[]{}
    \pgfplotstabletypeset[DAC improvements]{dac-compare-alpha-#2.csv}
  \end{table}
}

\dacCompareTable{Zero}{zero}{}
\dacCompareTable{Any}{any}{boosted by our machine-learning-assisted metric}

To compare the selected traditional techniques to DAC we use the
benchmark we introduced in RQ1 as a means of consistent evaluation. Both DAC and
our selection of traditional techniques require some number of hyper-parameters
to be set. To ensure a fair evaluation, we have searched over a range of
hyper-parameters for each of the selected techniques and compare between the
best configurations for each technique. Table~\ref{Ta:DACAlphaZero} provides performance
measurements for each of the three off-the-shelf clustering baselines across each of our five
target projects. In addition, \tableref{DACAlphaZero} provides the relative performance
increase gained by using DAC in place of these baselines.\footnote{We compute the relative performance
increase by comparing to the best overall off-the-shelf technique on a per-project basis.} For this comparison we have
made only the co-occurrence distance matrix available to our off-the-shelf baselines
as one of DAC's key insights was the importance of a machine-learning-assisted metric.
Table~\ref{Ta:DACAlphaAny} follows the same format but provides each off-the-shelf technique
access to the combined matrix DAC uses for clustering. In either case, we see that
DAC outperforms each of the baselines by a wide margin.

\subsection{RQ3: How does the choice of word vector learner and the choice of
sampling techniques affect the resulting clusters?\label{Se:Rq-SAndL}}

\Cref{rq:choices} seeks to understand the impact of two choices made in the
earlier portion of our toolchain: the choice of word vector learner and the
choice of trace sampling technique. For the choice of word vector learner we
argued that fastText with its utilization of sub-word information (in the form
of character level n-grams) would provide embeddings better suited to the task
of extracting clean clusters. We based this prediction on the observation, made by
many, that similarly named methods often have similar meaning. When it came to
the choice of trace sampling we sought to reduce the impact of a problem,
identified by~\citet{henkel_code_2018}, called \emph{prefix dominance}. To
address this issue of \emph{prefix dominance} in our specification mining
setting we introduced a trace sampling methodology termed Diversity Sampling.

To understand the interplay and effects of these choices we have evaluated the
\verb|ml4spec| toolchain in nine configurations. These nine configurations
are defined by two choices: a choice of word vector learner (either fastText~\citep{bojanowski_enriching_2017},
GloVe~\citep{pennington_glove:_2014}, or word2vec~\citep{mikolov_distributed_2013}) and a choice of trace sampling technique (either Diversity
Sampling, random sampling, or no sampling). By evaluating our full toolchain
with varying choices of embedding and sampling methodology we can either confirm
or refute our intuitions. We leverage the gold standard clusters
introduced in RQ1 to provide a consistent benchmark for comparison between the
nine configurations we've outlined. 

\newcommand{\samplerLearnerTable}[2]{
  \begin{table}
    \captionsetup{width=0.75\textwidth}
    \caption{Top-#1 performance (geometric mean across our five target projects). The \tikzemph{fill=best}{shaded row and column} represent the best sampler and learner, respectively.\label{Ta:sampler-by-learner-table-#2}}
    \Description[]{}
    \pgfplotstabletypeset[highlight best]{sampler-learner-#2-table.csv}
  \end{table}
}

\samplerLearnerTable{5}{ave}

First, we examine top-5 performance (measured against our benchmark) across
all of the configurations we established in~\sectref{GridSearch}. We look at
averages of the top-5 configurations (with sampler and learner fixed to one of
the nine choices outlined earlier) to understand effects of our
choices of interest (the sampler and learner) and to reduce any variance from
other sources. Table~\ref{Ta:sampler-by-learner-table-ave} provides top-5
performance measurements for each of our nine possible configurations.
We can see that fastText is superior (regardless of sampling
choice) to any of the other word vector learners by a wide margin. We also
observe that fastText paired with Diversity Sampling is the most performant
combination. However, fastText with no sampling is not far behind---this is
perhaps indicative of both the impact of word embeddings and the need for
larger corpora to learn suitable embeddings.

\samplerLearnerTable{1}{max}

Although top-5 averages provide a robust picture of the performance of our selected
configurations, we also would like to understand which configurations have the
best peak (top-1) performance. To assess top-1 performance, we examine
\tableref{sampler-by-learner-table-max} which shows the geometric mean
(across our five target projects) of the best performing configuration
identified via the data from our grid search (\sectref{GridSearch}).
These results affirm the impact of Diversity Sampling and fastText
as the superior word-embedding learner for this use case. Finally, we
observe that the combination of fastText and Diversity Sampling again produces
the best overall performance.

These results support two conclusions. First, fastText, with its use of sub-word
information, outperforms GloVe and word2vec in the cluster extraction task we
have benchmarked. Second, Diversity Sampling both improves the performance
of our toolchain and word vector learner (by reducing the amount of input data)
and provides an increase in performance compared to the other baseline choices
of sampling routine. These results also support further examination of the
advantages of sub-word information in the software-engineering domain;
specifically, we note that fastText has no concept of the ideal boundaries
between sub-tokens that naturally exist in program identifiers. A word vector
learner equipped with this knowledge may produce even more favorable results.

\subsection{RQ4: Is there a benefit to using a combination of co-occurrence statistics and word embeddings?\label{Se:Rq-Alpha}}

One of the key insights from~\sectref{DAC} was that word embeddings
and co-occurrence statistics capture subtly different
information. Word embeddings excel
at picking up on local context (a direct result of being based on the
distributional hypothesis which asserts that similar words appear in similar
contexts). This focus on local context makes word embeddings well-suited for
tasks like word similarity and analogy solving. For specification mining,
co-occurrence information is often used, in some form, to capture the
``support'' for a candidate rule or pattern. These co-occurrence statistics
encode a global relationship between words that is more far-reaching than the
relationship captured by word vectors.

\Cref{rq:benefit} attempts to precisely quantify the impact of these two
different sources of information. This effort is made somewhat easier by the
choice to include a tunable parameter in our toolchain that represents the
relative weight of word--word distance and co-occurrence distance in our final
pair-wise distance matrix. By evaluating our full toolchain with a gradation of
weight values we can pinpoint the mix of metrics that lead to optimal
performance on the benchmark we introduced earlier.

\newcommand{\varyingAlphaFigure}[3]{
  \begin{figure}
    \begin{tikzpicture}
      \begin{axis}[varying alpha=#3]
        \pgfplotstableread{alpha-plot-#2.csv}{\results}
        \foreach \benchmark in {redis,curl,hexchat,nmap,nginx} {
          \addplot table [y=\benchmark] from \results;
          \edef\temp{\noexpand\addlegendentry{\benchmark}}
          \temp
        }
      \end{axis}
    \end{tikzpicture}
    \caption{#1 benchmark performance for varying values of $\alpha$\label{Fi:alpha-plot-#2}}
    \Description[]{}
  \end{figure}
}

\varyingAlphaFigure{Average}{ave}{north}
\varyingAlphaFigure{Peak}{max}{south}

Figure~\ref{Fi:alpha-plot-ave} plots average benchmark scores for
different values of the $\alpha$ parameter. Here we take an average, with $\alpha$
fixed, over all configuration explored in~\sectref{GridSearch}. We observe a clear trend of
increasing performance as more weight is applied to the word embeddings (and
less to the co-occurrence statistics). In addition, we note that this performance
increase reaches a peak at $\alpha=0.75$ for each of the five projects. As we
did in \cref{rq:choices}, we also examine top-1 performance. The results for top-1 performance, given
in~\figref{alpha-plot-max}, paint a clearer picture of the relationship
between word embeddings and co-occurrence statistics. In~\figref{alpha-plot-max}
we observe that, for all five projects, adding word embeddings to our distance matrix
produces a pronounced increase in performance. Again we are able to validate that, for
each project, this increase in performance peaks at $\alpha=0.75$. These results suggests an affirmative
answer to \cref{rq:benefit}: there is a quantifiable benefit to using both
co-occurrence statistics and word embeddings; furthermore, a combination that favors
the distances produced via word embeddings yields maximum performance across each of the
projects we examined.

\subsection{RQ5: Does our toolchain transfer to unseen projects with minimal reconfiguration?\label{Se:Rq-Transfer}}

\Cref{rq:transfer} asked if our toolchain can be easily adapted to unseen
projects. More precisely, we would like to evaluate whether the few parameters
in our tool can use reasonable defaults without sacrificing too much performance. To
do this we can re-examine some of the results from the grid search in~\sectref{GridSearch} to develop an
understanding of the performance impact choosing defaults would have when
transferring to unseen data.

We can rank configurations by enumerating the top-20 configurations for each of our five target projects and
counting the number of times any given configuration occurs in the top-20 for any
project. This ranking reveals that two configurations
each produce top-20 results in four out of the five projects we considered. Examining
these two configurations further, we find that exploring just these two configurations
allows \verb|ml4spec| to find clusters that are within ten percent of the best performing
configuration for each of our five projects. Therefore, we can answer \cref{rq:transfer}
in the affirmative, with the knowledge that running in just two default configurations
allows for full automation with no more than a ten percent performance decrease.

% Now here all we need to do is look at that grid search across projects and
% look at the finer breakdown of parameters and see how far off the "best"
% parameters for redis are from the "best" parameters for hexchat, nmap, nginx,
% and curl

% Following this we could get into threats and related work 

% Finally, after all that, it would be great to have a future work and a
% conclusion that highlights yet another result in which we recover information
% captured in a recent work (APISan). This would be awesome because, although th
% comparison is decidedly not direct, the results provide confidence in the tool
% and make a strong statement about the scalability of this work.

% LocalWords:  Hexchat Ngnix Nmap Redis GridSearch RQ1 RQ2 DAC pre
% LocalWords:  DBSCAN RQ3 fastText GloVe word2vec RQ4 nginx RQ5 DAC's
% LocalWords:  GridSearchParams DACAlphaZero dac csv

\section{Threats to Validity\label{Se:Threats}}

Our usage of Parametric Lightweight Symbolic Execution (PLSE) provides us with
a way to quickly extract rich traces from buildable C programs; however,
PLSE is imprecise. It is possible that a more precise symbolic-execution engine would
provide our downstream mining techniques with more accurate information and, in turn,
reveal more correct specifications. In addition, it is likely that an execution engine
capable of generating interprocedural traces would improve the quality of our results.

Our ground-truth clustering benchmark was manually extracted with the goal of providing
a quantitative benchmark in the rich vocabulary available to the \verb|ml4spec| toolchain. This
manual process is susceptible to bias. To mitigate this risk, each cluster was validated against
the source program to confirm that the set of terms within the cluster appeared in a
concrete usage. Furthermore, the gold-standard clusters were created with no bounds on their size:
this variance in cluster size greatly increases the difficulty of recovering correct clusters while matching
the reality of usage patterns which can range from simple checks to complex iterators or initialization routines.
It is our hope that the release of the ground-truth dataset will provide the groundwork for a larger,
more comprehensive, gold-standard dataset curated by the community.

We chose to focus on a collection of five open-source C projects. It is possible that
our selection of projects is not representative of the wider landscape of API usages in C.
Furthermore, our technique, while language agnostic in theory, may not easily transfer
to other languages. Fortunately, the PLSE implementation uses the GCC toolchain as a
front end, which makes cross-language mining a possibility for future work.  

% LocalWords:  PLSE dataset

\section{Related Work\label{Se:RelatedWork}}

There exists a wide variety of related works from the specification mining,
API misuse, program understanding, and entity embedding communities. For comprehensive
overviews of specification mining and misuse we refer the reader to~\citet{lo_mining_2011} and~\citet{robillard_automated_2013}.
For efforts in machine learning and its application in the software-engineering domain~\citet{allamanis_survey_2017} provide
an excellent survey. In addition, there exists a listing of machine learning on code resources
maintained by the community~\citep{source_d_cool_2019}. For details on embeddings and their
use in the software-engineering domain~\citet{martin_monperrus_embeddings_2019} provide an up-to-date listing.
In the following sections, we discuss related works in the realms of specification mining
and embeddings of software artifacts in greater detail. 

\subsection*{Specification Mining}
There is a rich history of work on mining specifications, or usage patterns, from programs. Earlier approaches, such
as~\citet{li_pr-miner:_2005}, provided relatively simple specifications. Going
forward in time, a growing body of work attempted to produce richer FSA-based
specifications~\citep{lorenzoli_automatic_2008,ammons_mining_2002,pradel_automatic_2009,gabel_javert:_2008,walkinshaw_inferring_2008,walkinshaw_reverse_2007,quante_dynamic_2007,shoham_static_2008,dallmeier_mining_2006,acharya_mining_2009,lo_smartic:_2006}.
Some recent work such as Deep Specification Mining and Doc2Spec, has incorporated NLP techniques~\citep{zhong_inferring_2009,le_deep_2018}. \Citet{defreez_path-based_2018} use
word-vector learners to bolster traditional support-based mining via the identification of \emph{function synonyms}.
In the broader field of non-FSA-based specification mining techniques, there exist several novel
techniques: \citet{nguyen_graph-based_2009} mine graph-based specifications;
\citet{sankaranarayanan_mining_2008} produce specifications as Datalog rules;
\citet{acharya_mining_2007} create a partial order over function calls and \citet{Murali2017} develop
a Bayesian framework for learning probabilistic specifications. In addition to mining,
several works focus on the related problem of detecting misuses~\citep{Engler2001,yun_apisan:_2016,monperrus_detecting_2013,livshits_dynamine:_2005,wasylkowski_detecting_2007}.

The \verb|ml4spec| toolchain is agnostic to the choice of trace-based mining technique
used to generate specifications. This miner-agnostic perspective makes \verb|ml4spec|
a front end that enables prior trace-based miners to work in the Open-World setting we have
described. In addition, \verb|ml4spec|'s use of Parametric Lightweight Symbolic Execution
makes it possible to mine, via traditional methods, specifications that involve both
control-flow and data-flow information. 

\subsection*{Embeddings of Software Artifacts}

Recently, several techniques have leveraged learned embeddings for artifacts
generated from programs.
\Citet{Nguyen2017,Nguyen:2016:MAE:2889160.2892661} leverage word embeddings
(learned from ASTs) in two domains to facilitate translation from Java to C\#.
\Citet{le_deep_2018} use embeddings to bootstrap anomaly detection against a
corpus of JavaScript programs. \Citet{Gu:2016:DAL:2950290.2950334} leverage an
encoder/decoder architecture to embed whole sequences in their \textsc{DeepAPI}
tool for API recommendation.

\Citet{Pradel2017} use embeddings (learned from custom
tree-based contexts built from ASTs) to bootstrap anomaly detection against a
corpus of JavaScript programs.  \Citet{Gu:2016:DAL:2950290.2950334} leverage an
encoder/decoder architecture to embed whole sequences in their \textsc{DeepAPI}
tool for API recommendation. \textsc{API2API}
by \citet{7886921} also leverages word embeddings, but it learns the
embeddings from API-related natural-language documents instead of an artifact
derived directly from source code. \Citet{Alon:2018:GPR:3192366.3192412} learn
from paths through ASTs to produce general representations of programs; in \citep{DBLP:journals/corr/abs-1803-09473} 
they expand upon this general representation by leveraging attention mechanisms.
\Citet{ncc} produce embeddings of programs that are learned from both control-flow and 
data-flow information. \Citet{Zhao:2018:NSA:3236024.3236066} introduce type-directed encoders, a framework
for encoding compound data types via a recursive composition of more basic encoders. Using input/output pairs as the input data for learning,
\citet{Piech:2015:LPE:3045118.3045235} and \citet{DBLP:journals/corr/ParisottoMSLZK16} learn to embed whole
programs. Using sequences of live variable values, \citet{DBLP:journals/corr/abs-1711-07163}
produce embeddings to aid in program repair tasks. \Citet{DBLP:journals/corr/abs-1711-00740} learn to embed
whole programs via Gated Graph Recurrent Neural Networks (GG-RNNs)
\citep{DBLP:journals/corr/LiTBZ15}. \Citet{Peng:2015:BPV:2978872.2978936} provide 
an AST-based encoding of programs with the goal of facilitating
deep-learning methods.

In contrast to prior work on the embedding of software artifacts, we provide both
a novel use of embeddings in the software-engineering domain (in the form of Domain-Adapted Clustering and
its machine-learning-assisted metric) and a comprehensive comparison between three state-of-the-art word embedding techniques
(fastText~\citep{bojanowski_enriching_2017}, GloVe~\citep{pennington_glove:_2014}, and word2vec~\citep{mikolov_distributed_2013}).
Furthermore, we make an insight into a future line of work involving the utilization
of refined sub-token information to improve embeddings in the software-engineering domain.

% LocalWords:  Doc2Spec NLP DeepAPI API2API GG RNNs 's fastText GloVe
% LocalWords:  word2vec

% \input{supporting}
\section{Conclusion\label{Se:Conclusion}}

With a growing number of frameworks and libraries being authored each day,
there is an increased need for industrial-grade specification mining. In this
paper, we introduced the problem of Open-World Specification Mining with the
hope of fostering new mining tools and techniques that focus on reducing burdens
on users. The challenge of mining is amplified in an Open-World setting. To
address this challenge, we introduced \verb|ml4spec|: a toolchain
that combines the power of unsupervised learning (in the form of word embeddings) with
traditional techniques to successfully mine specifications and usage patterns in
an Open-World setting.

Our work also provides a new dataset of ground-truth clusters which can be used
to benchmark attempts to extract related terms from programs. We provided a comprehensive
evaluation across three different word-vector learners to gain insight into the value of
sub-word information in the software-engineering domain. Lastly, we introduced three
new techniques: Hierarchical Thresholding, Diversity Sampling, and Domain Adapted Clustering
each solving a different challenge in the realm of Open-World Specification Mining.

% \input{acknowledgments}

% \listofchanges{}

\bibliographystyle{acm/ACM-Reference-Format}
\bibliography{bib/henkel,bib/henkel-old}

\end{document}